\documentclass[]{agujournal}

\journalname{Geochemistry, Geophysics, Geosystems}
\usepackage[utf8]{inputenc}
\usepackage{graphicx}
\usepackage{bm}

\begin{document}

\title{Block Motion Changes in Japan Triggered by the 2011 Great Tohoku Earthquake}

\authors{Brendan J. Meade\affil{1}, John P. Loveless\affil{2}}
\affiliation{1}{Department of Earth and Planetary Sciences, Harvard University, Cambridge, Massachusetts, 02138}
\affiliation{2}{Department of Geosciences, Smith College, Northampton, Massachusetts, 01063}
\correspondingauthor{B. J. Meade}{meade@fas.harvard.edu}

\begin{keypoints}
\item Analytic theory for block motion changes from earthquakes
\item GPS constrained block motions reveal select block motion changes before and afer the 2011 Tohoku earthquake
\end{keypoints}

\begin{abstract}
Plate motions are governed by equilibrium between basal and edge forces. Great earthquakes may induce differential static stress changes across tectonic plates, enabling a new equilibrium state. Here we consider the torque balance for idealized circular plates and find a simple scalar relationship for changes in relative plate speed as a function of its size, upper mantle viscosity, and coseismic stress changes. Applied to Japan, the 2011 $\mathrm{M}_{\mathrm{W}}=9.0$ Tohoku earthquake generated coseismic stresses of $10^2-10^5$~Pa that could have induced changes in motion of small (radius $\sim100$~km) crustal blocks within Honshu. Analysis of time-dependent GPS velocities, with corrections for earthquake cycle effects, reveals that plate speeds may have changed by up to $\sim3$ mm/yr between $\sim3.75$-year epochs bracketing this earthquake, consistent with an upper mantle viscosity of $\sim 5\times10^{18}$Pa$\cdot$s, suggesting that great earthquakes may modulate motions of proximal crustal blocks at frequencies as high as $10^-8$~Hz.
\end{abstract}

\section{Introduction}
Plate tectonic theory provides a kinematic framework for describing crustal motion at the Earth's surface. Rates of motion for the largest tectonic plates have been inferred independently over million-year time-scales from geologic observations \citep{demets1990current} and decadal time-scales from geodetic measurements \citep{sella2002revel}. In many cases late-Cenozoic and decadal plate motion estimates agree to within 5\% \citep{demets1999new}, however $\sim$1 million year reorganization events have also been observed \citep{croon2008revised}. At smaller spatial scales ($1,000$ km) and shorter temporal scales ($<10,000$ years), earthquake and fault activity at some plate boundaries has also been found to be irregular in time \citep{bennett2004codependent, dolan2007long, dolan2016extreme}. A hypothesis that may provide an explanation for such short-term variations is that plate motions may be perturbed over short time scales due to changes in the crustal and upper mantle stress state caused by great earthquakes \citep{anderson1975accelerated, romanowicz1993spatiotemporal} or evolving earthquake sequences \citep{ismail1999numerical, ismail2007numerical}. A geodetically based inference of accelerated Pacific plate subduction following the 2011 Tohoku, Japan earthquake may provide an example of the direct observation of this type of behavior \citep{heki2013accelerated}. Though this interpretation is non-unique \citep{tomita2015first} it is consistent with the inference of accelerated subduction rates beneath Kanto inferred from repeating earthquakes \citep{uchida2016acceleration}. However, because stress changes are largest in the immediate vicinity of a great earthquake, possible changes in motion are likely to be observed in crustal blocks adjacent to the source. In particular, small tectonic plates or crustal blocks in the hanging wall above subduction zones may experience relatively large coseismic stress changes ($10^2-10^5$~Pa) along their edges with relatively little resistance to motion at the crust-mantle interface due to the fact that viscosities may be relatively low in subduction zone mantle wedges \citep{billen2001low, kelemen2003thermal}, resulting in a coseismically induced change in plate motion.

Here we consider this argument quantitatively by calculating a first-order approximation of the magnitude of the crustal block motion change induced by coseismic stress changes using a simple disk geometry for crustal plates and assuming a Newtonian mantle rheology. These first-order predictions can be tested against geodetically constrained estimates of velocity changes before and after both the 2003 $\mathrm{M}_{\mathrm{W}}=8.3$ Tokachi and 2011 $\mathrm{M}_{\mathrm{W}}=9.0$ Tohoku earthquakes in Japan.

A challenge to the direct observation of such motion changes near major plate boundaries is that geodetically observed surface motions reflect the combined, and correlated, contributions from both plate motions and earthquake cycle processes. In other words, interseismic strain accumulation, postseismic deformation, and plate rotations must be disentangled in the vicinity of large earthquakes. Here we utilize results from time-dependent block models of Japan that simultaneously consider these contributions in a way that minimizes temporal variations in nominally interseismic station velocity changes \citep{loveless2016two}.

\section{Coseismic Stress and Plate Motion Changes}
The motions of tectonic plates are in quasi-static force equilibrium if there is no angular acceleration and the sum of applied torques equals zero. The quasi-static torques may be partitioned into edge, $\mathbf{T}_\mathrm{E}$, and basal, $\mathbf{T}_\mathrm{B}$, components \citep{solomon1974some, forsyth1975relative, iaffaldano2015rapid} as,

\begin{equation}
\int \mathbf{T}_\mathrm{E} da_\mathrm{E} + \int \mathbf{T}_\mathrm{B} da_\mathrm{B} = 0,
\end{equation}

with $\mathbf{T}_\mathrm{E} = \mathbf{x} \times (\boldsymbol{\sigma}_\mathrm{E} \cdot \mathbf{n}_\mathrm{E})$ and $\mathbf{T}_\mathrm{B} = \mathbf{x} \times (\boldsymbol{\sigma}_\mathrm{B} \cdot \mathbf{n}_\mathrm{B})$. Here $\mathbf{x}$ is position, $\boldsymbol{\sigma}_\mathrm{E}$ is the stress at the seismogenic edges of the block, $\boldsymbol{\sigma}_\mathrm{B}$ is the viscous stress exerted by the mantle on the base of the block, $\mathbf{n}_\mathrm{E}$ is a unit vector normal to the fault at the block edge, and $\mathbf{n}_\mathrm{B}$ is a unit vector normal to the fault at the block base. The viscous stresses for an iso-viscous Newtonian mantle are $\boldsymbol{\sigma}_\mathrm{B} = \eta \boldsymbol{\epsilon}_\mathrm{B}$, where, $\eta$ is the viscosity of the mantle and $\boldsymbol{\epsilon}_\mathrm{B}$ is the strain rate at the base of the tectonic plate. We calculate the relative magnitudes of these torques considering the case of a plate with the shape of circular disk with radius $r$ and bounding faults extending from the surface to depth, $h$. A first-order approximation of the magnitude of the edge torque, $T_\mathrm{E}$, is given by,

\begin{equation}
T_\mathrm{E} = \left| \left| \int \mathbf{x} \times (\bm{\sigma}_\mathrm{E} \cdot \mathbf{n}_\mathrm{E}) da_\mathrm{E} \right| \right| = R \| \bm{\sigma}_\mathrm{E}\| A_\mathrm{E} = 2\pi R r h \sigma_\mathrm{E},
\end{equation}

where $R$ is the radius of the Earth, the area of the plate edge is $A_\mathrm{E}=2\pi r h$, and $\| \bm{\sigma}_\mathrm{E}\|$ gives the average stress on the edges of the plate. Similarly, a first-order approximation of the magnitude of the basal torque, $T_\mathrm{B}$, is given by,

\begin{equation}
T_\mathrm{B} = \left| \left| \int \mathbf{x} \times (\boldsymbol{\sigma}_\mathrm{B} \cdot \mathbf{n}_\mathrm{B}) da_\mathrm{B} \right| \right| = R \eta \| \dot{\boldsymbol{\epsilon}}_\mathrm{B}\| A_\mathrm{B} = \frac{\pi R r^2 \eta s}{D},
\end{equation}

where the basal area of the plate is $A_\mathrm{B} = \pi r^2$ and the strain rate in the upper mantle, $\dot{\boldsymbol{\epsilon}}_\mathrm{B}$, is given by the plate speed, $s$, divided by the thickness of an upper mantle-like layer, $D$, in the Couette flow approximation. Because we are concerned with the case where the quasi-static torque balance is satisfied, both before and after a coseismic stress change, we assume that any time variation is balanced and plate geometry fixed,

\begin{equation}
\frac{d}{dt} \left( T_\mathrm{E} + T_\mathrm{B} \right) = 2h \frac{d\sigma_\mathrm{E}}{dt} + \frac{\eta r}{D} \frac{ds}{dt}= 0.
\end{equation}

\begin{figure}
\centering
\includegraphics[width=1.0\textwidth]{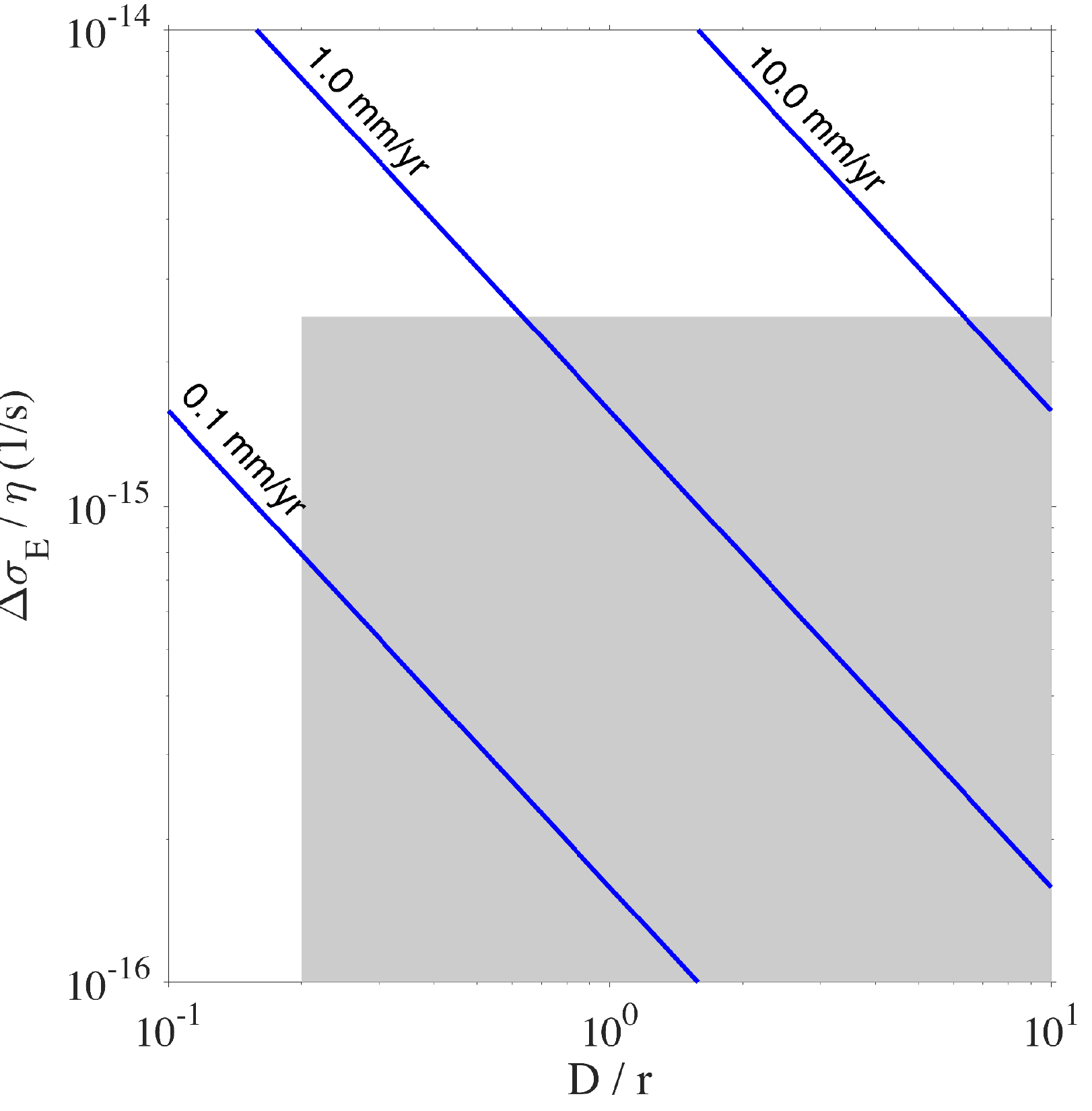}
\caption{Blue lines show predicted changes in speed, $\Delta s$, of crustal blocks as afunction of coseismic stress change, $\Delta \sigma_\mathrm{E}$, block radius, $r$, depth to fixed mantle, $D$, and upper mantle viscosity, $\eta$. Upper plate thickness is assumed to be $h=15$~km.	Over the parameter ranges applicable to Japan (gray region), block speed changes range from $0-10$~mm/yr.}
\label{PredictedSpeedPerturbations}
\end{figure}

Discretizing over finite changes in coseismic stress, $d\sigma_\mathrm{E} / dt \approx \Delta \sigma_\mathrm{E} / \Delta t$, and plate speed, $ds / dt \approx \Delta s / \Delta t$, the change in plate speed as a function of coseismic stress changes is,

\begin{equation}
\Delta s = \frac{2 h D \Delta \sigma_\mathrm{E}}{\eta r},
\end{equation}

which may be interpreted as the change in plate speed, without regard to change in direction, as a result of coseismic stress changes in the upper crust. In this idealized quasi-static formulation the change in plate velocity occurs over seismic wave propagation time scales, temporally coincident with the change in coseismic stress.  We can estimate the potential magnitude of this effect with characteristic parameters, $h = 15$ km, $D=100$ km. Upper mantle viscosities beneath Japan have been estimated independently from studies of both post-glacial and post-seismic rebound and range from $4\times10^{18}$ to $5\times10^{20}$~Pa$\cdot$s (Table~\ref{table1}) \citep{nur1974postseismic, nakada1986holocene, nakada1991late, suito1999simulation, ueda2003postseismic, nyst20061923, mavrommatis2014decadal}. In general, viscosity estimates at shorter forcing frequencies (100--1,000 year earthquake cycles) are $\sim 10\times$ lower than those inferred at longer duration forcing frequencies ($\sim100,000$ year glacial cycles). For intermediate field stress drops ($10^2-10^5$Pa) associated with the Tohoku earthquake \citep{toda2011using} and crustal blocks of radius 30 km, we estimate coseismically induced plate speed changes of ranging from $0.05-5.0$ mm/yr (Figure \ref{PredictedSpeedPerturbations}).  A weakness of this scalar approach is that it only reflects changes in plate speed relative to a fixed point in the mantle. What this means is that there are cases where changes in plate velocity would not be captured with this theory. For example, if a plate were moving at $5$ mm/yr to the north before an earthquake and then $5$ mm/yr to the south after the earthquake, $\Delta s=0$, yet the change in plate velocity is $10$ mm/yr.  The full tensor version of the theory addresses this issue but is beyond the scope of this paper.

\section{Block Speed Changes in Japan}
Due to the fact that GPS observations record not only plate motions but also earthquake cycle processes, the detection of plate motion changes requires an estimate of the relative contributions of these two effects. Previous work on the decomposition and block modeling of $~\sim19$~years of time-dependent GPS positions in Japan \citep{loveless2016two} has led to the development of minimally variable velocity fields over five temporally distinct  year long epochs (Table~\ref{table2}). The multi-year duration of these epochs is longer than the sub-hourly time scale involved in the idealized quasi-static plate motion change theory described above. The reason for this is that these epochs are bracketed by specific seismic events, enabling comparison with coseismically imposed stress change, and are long enough that they can be decomposed into constitutive parts. In this work we assumed that, over each epoch, the GPS position time series at each station can be decomposed into four contributions: 1) linear trend (epoch velocity), 2) step function (coseismic displacements and equipment maintenance), 3) harmonic terms (annual and bi-annual signals), and 4) two exponentially decaying terms (accounting for short- and long-term post-seismic deformation). Epoch velocities estimated over these intervals are determined in way that is minimally variable in the sense that the relaxation timescale for the longer exponential term is determined so that the linear terms, which we take to represent nominally interseismic velocity fields, are as similar as possible from one epoch to the next.

These interseismic velocity fields are used as inputs into block models that simultaneously estimate block motions and earthquake cycle effects due to coupling on block-bounding faults and the subduction interfaces off the Pacific coast of Japan.  It is the velocity component arising from block motions, rather than the raw GPS velocities, that we consider here for the identification of changes in plate speeds, $\Delta s$. In other words, changes in plate motions are determined by considering epoch-to-epoch differences in the estimates of block rotations derived from block model decomposition of the minimally variable GPS velocity fields.  These changes in estimated rotation vectors (Euler poles) form the basis for this analysis.  Instead of describing the change in motion of each block with a single scalar speed we consider a frequency distribution of velocities by calculating differential velocities at 1,000 randomly sampled geographic locations on each block (Figure~\ref{JapanDeltaS}). This approach allows us to simply see spatially complex behaviors such as those arising from local Euler poles, which are manifest as frequency distributions with large spreads (e.g., $>6$ mm/yr, Figure \ref{JapanDeltaS}).

Differential epoch-to-epoch block speeds are given by, $\Delta s_i = | \langle s(E_{i+1}) \rangle - \langle s(E_{i}) \rangle |$, where $\langle s(E_{i}) \rangle$ is the mean of the frequency distribution of location-by-location velocity during the epoch $E_i$  (Table~\ref{table2}). To determine if there are distinct changes in block motions associated with the geodetic observation epochs bracketing the 2003 Tokachi and 2011 Tohoku earthquakes, we define a block speed change index. This index evaluates as true for a given block if all three speed change conditions are met. First, the mean epoch-to-epoch speed change is greater than 5~mm/yr, $\langle \Delta s_i \rangle > 5$. This criterion identifies only the largest plate motion changes. Second, the variability of differential speeds calculated at points on a given block is less than 4~mm/yr, i.e., $SD(\Delta s_i) < 4$. These criteria limit the misidentification of block motion changes that are due to local rotation rate changes.  We exclude these cases because the scalar theory only represents changes in plate speed. Third, the average epoch-to-epoch variability is less than 4~mm/yr, i.e., $\langle SD(\Delta s_i) < 4 \rangle$. This criterion precludes the misidentification of velocity changes in noisy regions. These three criteria are evaluated for each of the upper plate blocks in Japan.

Using these motion change detection criteria, we find no significant epoch-to-epoch speed changes other than between the intervals spanning the 2011 Tohoku earthquake.  This includes a lack of identifiable plate speed changes in the epochs bracketing the 2003 Tokachi earthquake, which may be explained by the facts that there appear to be no small crustal blocks in the immediate vicinity of this event (Hokkaido and northern Honshu) where the coseismic stress changes were largest, and the magnitude of stress change was smaller than that due to the 2011 event. In contrast, we detect four blocks that appear to have discernable changes in plate speed in the epoch bracketing the 2011 Tohoku earthquake (Figure~\ref{JapanDeltaS}). Each of these blocks lies in central Honshu near the Itoigawa-Shizuoka tectonic line and Niigata-Kobe tectonic zone and has a characteristic length scale of $\sim 100$~km. The differential speed estimates for these blocks range from $2-5$~mm/yr, assuming upper mantle viscosities $>10^{18}$~Pa$\cdot$s (Figure~\ref{PredictedSpeedPerturbations}). Note that the change detection index does not identify plate motion changes in the block immediately above the rupture area of the Tohoku earthquake in eastern Honshu where geodetic observations suggest that postseismic deformation is greatest \citep{ozawa2011coseismic}.

\section{Discussion}
Slip on geometrically complex fault systems enables tectonic plates to move past one another. At these interfaces, fault system geometry may evolve in response to geometric \citep{gabrielov1996geometric} and kinematic \citep{mckenzie1969evolution} inconsistencies generated by repeated fault slip and cumulative fault offset. Similarly, fault slip rates are governed by the quasistatic equilibrium that sets the differential motion of bounding tectonic blocks. At longer time scales ($>1$~million year), changes in plate motion have been ascribed to changes in the relative buoyancy of subducted geometry/material as well as the collision and growth of tectonic terranes \citep{iaffaldano2006feedback, jagoutz2015anomalously}.  At $1,000-100,000$ year time scales, emerging geological evidence suggests that rates of local fault system activity may vary significantly ($3-10\times$) in both strike-slip and thrust faulting environments \citep{dolan2016extreme, saint2016major}.  Here we consider the possibility that short-term changes in fault slip rates may be the result of short-term changes in plate motions caused by coseismic stress changes. In other words, these observable changes in fault slip rates may be due to changes in the differential motion of bounding tectonic plates. This hypothesis provides an intrinsic source of variations in rates of fault system activity, with the implication that coseismically induced block motion changes persist sufficiently long to be captured in geologic records.

Here we’ve considered the case where coseismic stress changes modulate plate motions.  However, this is an inherently incomplete description of potential changes in plate motions even in the narrow context earthquake cycle behaviors. For example, the time-varying effects of viscoelastic post-seismic deformation may spread over 100s--1000s of km \citep{pollitz1998viscosity}.  Time-dependent viscous deformation of the upper mantle could therefore cause accelerated or decelerated surface velocities across an entire plate, blurring the conceptual distinction between classes of models other than the fact that the viscoelastic effect would decay with time and the coseismic stress-based model discussed here would be a more discrete change in plate motions. The viscoelastic explanation appears to be most in line with the previous suggestions of changing tectonic plate motions \citep{anderson1975accelerated, romanowicz1993spatiotemporal, pollitz1998viscosity} while the latter explanation is most consistent with ideas about fault and block systems \citep{gabrielov1996geometric, ismail1999numerical}.  An additional aliasing is possible when considering both coseismic displacements and changes in block motion to be step-like in time. This may be considered a simplification of the idealized version of the theory presented here and a more temporally complex type of behavior may be plausible where evolving stresses throughout the earthquake cycle modulate block and plate motions more gradually.

For the Japan example considered here, the epoch-wise velocity fields used in the block models \citep{loveless2016two} were estimated after subtracting out annual and semi-annual harmonics, coseismic displacements of up to 200 earthquakes, and two exponential terms (short- and long-duration) following seven major earthquakes during the time span considered.  If these exponentially decaying terms are interpreted as representing all postseismic deformation then the block motion changes estimated here can be interpreted as uniquely distinguished from crustal deformation that is driven by viscoelastic or afterslip processes.  Similarly, the observation that the block motion change index used here does not identify the large blocks closest to the 2011 Tohoku earthquake may indicate that there is minimal aliasing between changes in block motion and more classical postseismic deformation processes. However, there is significant covariance in velocity field decomposition between the estimated linear velocity and duration of postseismic deformation even when temporal variations in the nominally interseismic velocity term are minimized \citep{loveless2016two}.  The use of such velocity fields in the coseismic block motion change theory described here may therefore be considered to represent a parsimonious end-member interpretation.

\begin{figure}
\centering
\includegraphics[width=1.0\textwidth]{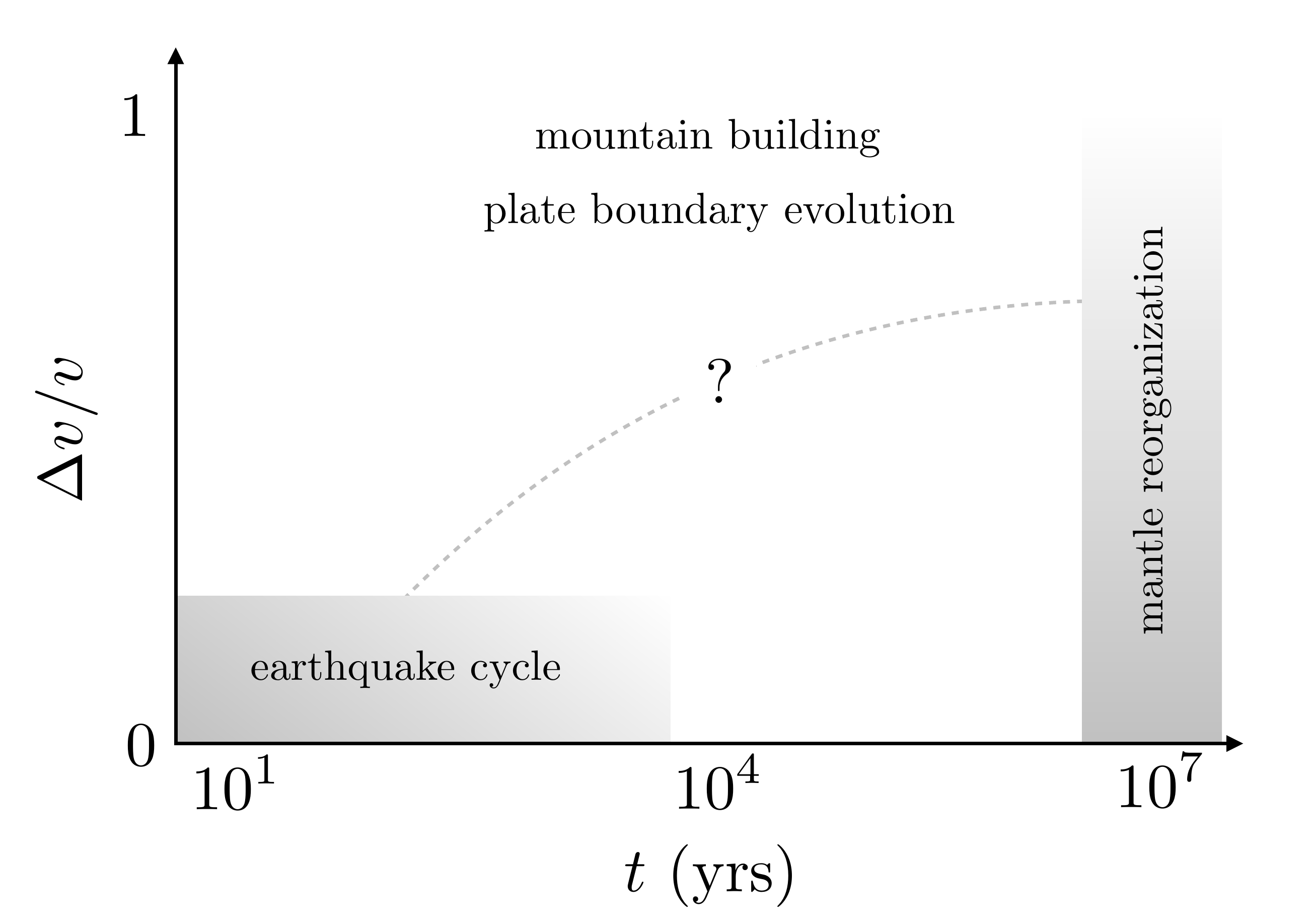}
\caption{Schematic diagram of sources of plate motion variations, $\Delta v / v$, as a function of time, $t$. Shaded regions represent constraints from the earthquake cycle component described in this paper. The short time scale variations suggested in this paper are at high-frequencies, $10^{-8}$~Hz, relative to mantle reorganization frequencies, $10^{-15}$~Hz.}
\label{PlateVelocityTimeSeriesFigure}
\end{figure}

In addition to this solid-earth focused hypothesis for rapid plate and fault motion evolution, climatically driven changes in surface loading have also been suggested to modulate fault slip rates \citep{hetzel2005slip, luttrell2007modulation}. Taken together there appears to be a growing suite of mechanisms that can modulate the rates of fault system behavior over relatively short time intervals including not only slip rate variations but also changes in interseismic loading rates \citep{sieh2008earthquake}. These processes may contribute to understanding of short-term fault and plate motion rate changes as well as contribute to an emerging view of tectonic motions as noisy trends through time (Figure~\ref{PlateVelocityTimeSeriesFigure}), where short-term (as low as $10^{-8}$~Hz) variations in fault system activity due to earthquakes and changes in surface loading lead to perturbations to long-term behavior. Whether or not a cluster of large earthquakes in a given region might change plate motions constructively or destructively over a large area remains an extant question.

\section{Conclusions}
The equilibrium torque balance on tectonic plates at short length scales ($<500$~km) suggests that plate speeds may be moderately perturbed ($<5$~mm/yr) by large coseismic stress changes ($10^3$~Pa) if the mantle beneath the crust is characterized by a relatively low viscosity ($<10^{19}$~Pa$\cdot$s). Following the 2011 $\mathrm{M}_\mathrm{W} = 9.0$ Tohoku earthquake in Japan we identify possible motion changes ($2-4$~mm/yr) in four small upper plate blocks across central and southern Honshu. Similar effects are not observed in blocks that are larger and/or more distant from the rupture source, nor following the smaller 2003  $\mathrm{M}_\mathrm{W} = 8.3$ Tokachi earthquake. This coseismic perturbation to block motion contributes to growing list of candidate processes that may modulate the temporal evolution of block and fault system activity at plate boundaries.

\begin{acknowledgments}
TBD
\end{acknowledgments}

\bibliographystyle{agufull08}
\bibliography{biblio}

\begin{thebibliography}{38}
\providecommand{\natexlab}[1]{#1}
\expandafter\ifx\csname urlstyle\endcsname\relax
  \providecommand{\doi}[1]{doi:\discretionary{}{}{}#1}\else
  \providecommand{\doi}{doi:\discretionary{}{}{}\begingroup
  \urlstyle{rm}\Url}\fi

\bibitem[{\textit{Anderson}(1975)}]{anderson1975accelerated}
Anderson, D.~L. (1975), Accelerated plate tectonics, \textit{Science},
  \textit{187}(4181), 1077--1079.

\bibitem[{\textit{Bennett et~al.}(2004)\textit{Bennett, Friedrich, and
  Furlong}}]{bennett2004codependent}
Bennett, R.~A., A.~M. Friedrich, and K.~P. Furlong (2004), {Codependent
  histories of the San Andreas and San Jacinto fault zones from inversion of
  fault displacement rates}, \textit{Geology}, \textit{32}(11), 961--964.

\bibitem[{\textit{Billen and Gurnis}(2001)}]{billen2001low}
Billen, M.~I., and M.~Gurnis (2001), {A low viscosity wedge in subduction
  zones}, \textit{Earth and Planetary Science Letters}, \textit{193}(1),
  227--236.

\bibitem[{\textit{Croon et~al.}(2008)\textit{Croon, Cande, and
  Stock}}]{croon2008revised}
Croon, M.~B., S.~C. Cande, and J.~M. Stock (2008), {Revised Pacific-Antarctic
  plate motions and geophysics of the Menard Fracture Zone},
  \textit{Geochemistry, Geophysics, Geosystems}, \textit{9}(7).

\bibitem[{\textit{DeMets and Dixon}(1999)}]{demets1999new}
DeMets, C., and T.~H. Dixon (1999), {New kinematic models for Pacific-North
  America motion from 3 Ma to present, I: Evidence for steady motion and biases
  in the NUVEL-1A Model}, \textit{Geophysical Research Letters},
  \textit{26}(13), 1921--1924.

\bibitem[{\textit{DeMets et~al.}(1990)\textit{DeMets, Gordon, Argus, and
  Stein}}]{demets1990current}
DeMets, C., R.~G. Gordon, D.~Argus, and S.~Stein (1990), {Current plate
  motions}, \textit{Geophysical Journal International}, \textit{101}(2),
  425--478.

\bibitem[{\textit{Dolan et~al.}(2007)\textit{Dolan, Bowman, and
  Sammis}}]{dolan2007long}
Dolan, J.~F., D.~D. Bowman, and C.~G. Sammis (2007), {Long-range and long-term
  fault interactions in Southern California}, \textit{Geology}, \textit{35}(9),
  855--858.

\bibitem[{\textit{Dolan et~al.}(2016)\textit{Dolan, McAuliffe, Rhodes, McGill,
  and Zinke}}]{dolan2016extreme}
Dolan, J.~F., L.~J. McAuliffe, E.~J. Rhodes, S.~F. McGill, and R.~Zinke (2016),
  {Extreme multi-millennial slip rate variations on the Garlock fault,
  California: Strain super-cycles, potentially time-variable fault strength,
  and implications for system-level earthquake occurrence}, \textit{Earth and
  Planetary Science Letters}, \textit{446}, 123--136.

\bibitem[{\textit{Forsyth and Uyeda}(1975)}]{forsyth1975relative}
Forsyth, D., and S.~Uyeda (1975), {On the relative importance of the driving
  forces of plate motion}, \textit{Geophysical Journal International},
  \textit{43}(1), 163--200.

\bibitem[{\textit{Gabrielov et~al.}(1996)\textit{Gabrielov, Keilis-Borok, and
  Jackson}}]{gabrielov1996geometric}
Gabrielov, A., V.~Keilis-Borok, and D.~D. Jackson (1996), {Geometric
  incompatibility in a fault system}, \textit{Proceedings of the National
  Academy of Sciences}, \textit{93}(9), 3838--3842.

\bibitem[{\textit{Heki and Mitsui}(2013)}]{heki2013accelerated}
Heki, K., and Y.~Mitsui (2013), {Accelerated Pacific plate subduction following
  interplate thrust earthquakes at the Japan trench}, \textit{Earth and
  Planetary Science Letters}, \textit{363}, 44--49.

\bibitem[{\textit{Hetzel and Hampel}(2005)}]{hetzel2005slip}
Hetzel, R., and A.~Hampel (2005), {Slip rate variations on normal faults during
  glacial--interglacial changes in surface loads}, \textit{Nature},
  \textit{435}(7038), 81--84.

\bibitem[{\textit{Iaffaldano and Bunge}(2015)}]{iaffaldano2015rapid}
Iaffaldano, G., and H.-P. Bunge (2015), {Rapid plate motion variations through
  geological time: Observations serving geodynamic interpretation},
  \textit{Annual Review of Earth and Planetary Sciences}, \textit{43},
  571--592.

\bibitem[{\textit{Iaffaldano et~al.}(2006)\textit{Iaffaldano, Bunge, and
  Dixon}}]{iaffaldano2006feedback}
Iaffaldano, G., H.-P. Bunge, and T.~H. Dixon (2006), {Feedback between mountain
  belt growth and plate convergence}, \textit{Geology}, \textit{34}(10),
  893--896.

\bibitem[{\textit{Ismail-Zadeh et~al.}(1999)\textit{Ismail-Zadeh, Keilis-Borok,
  and Soloviev}}]{ismail1999numerical}
Ismail-Zadeh, A., V.~I. Keilis-Borok, and A.~A. Soloviev (1999), {Numerical
  modelling of earthquake flow in the southeastern Carpathians (Vrancea):
  effect of a sinking slab}, \textit{Physics of the Earth and Planetary
  Interiors}, \textit{111}(3), 267--274.

\bibitem[{\textit{Ismail-Zadeh et~al.}(2007)\textit{Ismail-Zadeh, Le~Mou{\"e}l,
  Soloviev, Tapponnier, and Vorovieva}}]{ismail2007numerical}
Ismail-Zadeh, A., J.-L. Le~Mou{\"e}l, A.~Soloviev, P.~Tapponnier, and
  I.~Vorovieva (2007), {Numerical modeling of crustal block-and-fault dynamics,
  earthquakes and slip rates in the Tibet-Himalayan region}, \textit{Earth and
  Planetary Science Letters}, \textit{258}(3), 465--485.

\bibitem[{\textit{Jagoutz et~al.}(2015)\textit{Jagoutz, Royden, Holt, and
  Becker}}]{jagoutz2015anomalously}
Jagoutz, O., L.~Royden, A.~F. Holt, and T.~W. Becker (2015), {Anomalously fast
  convergence of India and Eurasia caused by double subduction}, \textit{Nature
  Geoscience}, \textit{8}(6), 475--478.

\bibitem[{\textit{Kelemen et~al.}(2003)\textit{Kelemen, Rilling, Parmentier,
  Mehl, and Hacker}}]{kelemen2003thermal}
Kelemen, P.~B., J.~L. Rilling, E.~Parmentier, L.~Mehl, and B.~R. Hacker (2003),
  {Thermal structure due to solid-state flow in the mantle wedge beneath arcs},
  \textit{Inside the Subduction Factory}, pp. 293--311.

\bibitem[{\textit{Loveless and Meade}(2016)}]{loveless2016two}
Loveless, J.~P., and B.~J. Meade (2016), {Two decades of spatiotemporal
  variations in subduction zone coupling offshore Japan}, \textit{Earth and
  Planetary Science Letters}, \textit{436}, 19--30.

\bibitem[{\textit{Luttrell et~al.}(2007)\textit{Luttrell, Sandwell,
  Smith-Konter, Bills, and Bock}}]{luttrell2007modulation}
Luttrell, K., D.~Sandwell, B.~Smith-Konter, B.~Bills, and Y.~Bock (2007),
  {Modulation of the earthquake cycle at the southern San Andreas fault by lake
  loading}, \textit{Journal of Geophysical Research}, \textit{112}(B8).

\bibitem[{\textit{Mavrommatis et~al.}(2014)\textit{Mavrommatis, Segall, and
  Johnson}}]{mavrommatis2014decadal}
Mavrommatis, A.~P., P.~Segall, and K.~M. Johnson (2014), {A decadal-scale
  deformation transient prior to the 2011 Mw 9.0 Tohoku-oki earthquake},
  \textit{Geophysical Research Letters}, \textit{41}(13), 4486--4494.

\bibitem[{\textit{McKenzie and Morgan}(1969)}]{mckenzie1969evolution}
McKenzie, D.~P., and W.~Morgan (1969), {Evolution of triple junctions},
  \textit{Nature}, \textit{224}(5215), 125--133.

\bibitem[{\textit{Nakada}(1986)}]{nakada1986holocene}
Nakada, M. (1986), {Holocene sea levels in oceanic islands: implications for
  the rheological structure of the Earth's mantle}, \textit{Tectonophysics},
  \textit{121}(2-4), 263--276.

\bibitem[{\textit{Nakada et~al.}(1991)\textit{Nakada, Yonekura, and
  Lambeck}}]{nakada1991late}
Nakada, M., N.~Yonekura, and K.~Lambeck (1991), {Late Pleistocene and Halocene
  sea-level changes in Japan: implications for tectonic histories and mantle
  rheology}, \textit{Palaeogeography, Palaeoclimatology, Palaeoecology},
  \textit{85}(1-2), 107--122.

\bibitem[{\textit{Nur and Mavko}(1974)}]{nur1974postseismic}
Nur, A., and G.~Mavko (1974), {Postseismic viscoelastic rebound},
  \textit{Science}, \textit{183}(4121), 204--206.

\bibitem[{\textit{Nyst et~al.}(2006)\textit{Nyst, Nishimura, Pollitz, and
  Thatcher}}]{nyst20061923}
Nyst, M., T.~Nishimura, F.~Pollitz, and W.~Thatcher (2006), {The 1923 Kanto
  earthquake reevaluated using a newly augmented geodetic data set},
  \textit{Journal of Geophysical Research}, \textit{111}(B11).

\bibitem[{\textit{Ozawa et~al.}(2011)\textit{Ozawa, Nishimura, Suito,
  Kobayashi, Tobita, and Imakiire}}]{ozawa2011coseismic}
Ozawa, S., T.~Nishimura, H.~Suito, T.~Kobayashi, M.~Tobita, and T.~Imakiire
  (2011), {Coseismic and postseismic slip of the 2011 magnitude-9 Tohoku-Oki
  earthquake}, \textit{Nature}, \textit{475}(7356), 373--376.

\bibitem[{\textit{Pollitz et~al.}(1998)\textit{Pollitz, B{\"u}rgmann, and
  Romanowicz}}]{pollitz1998viscosity}
Pollitz, F.~F., R.~B{\"u}rgmann, and B.~Romanowicz (1998), {Viscosity of
  oceanic asthenosphere inferred from remote triggering of earthquakes},
  \textit{Science}, \textit{280}(5367), 1245--1249.

\bibitem[{\textit{Romanowicz}(1993)}]{romanowicz1993spatiotemporal}
Romanowicz, B. (1993), {Spatiotemporal patterns in the energy release of great
  earthquakes}, \textit{Science}, \textit{260}(5116), 1923--1927.

\bibitem[{\textit{Saint-Carlier et~al.}(2016)\textit{Saint-Carlier, Charreau,
  Lav{\'e}, Blard, Dominguez, Avouac, Wang, Team et~al.}}]{saint2016major}
Saint-Carlier, D., J.~Charreau, J.~Lav{\'e}, P.-H. Blard, S.~Dominguez, J.-P.
  Avouac, S.~Wang, A.~Team, et~al. (2016), {Major temporal variations in
  shortening rate absorbed along a large active fold of the southeastern
  Tianshan piedmont (China)}, \textit{Earth and Planetary Science Letters},
  \textit{434}, 333--348.

\bibitem[{\textit{Sella et~al.}(2002)\textit{Sella, Dixon, and
  Mao}}]{sella2002revel}
Sella, G.~F., T.~H. Dixon, and A.~Mao (2002), {REVEL: A model for recent plate
  velocities from space geodesy}, \textit{Journal of Geophysical Research},
  \textit{107}(B4).

\bibitem[{\textit{Sieh et~al.}(2008)\textit{Sieh, Natawidjaja, Meltzner, Shen,
  Cheng, Li, Suwargadi, Galetzka, Philibosian, and
  Edwards}}]{sieh2008earthquake}
Sieh, K., D.~H. Natawidjaja, A.~J. Meltzner, C.-C. Shen, H.~Cheng, K.-S. Li,
  B.~W. Suwargadi, J.~Galetzka, B.~Philibosian, and R.~L. Edwards (2008),
  {Earthquake supercycles inferred from sea-level changes recorded in the
  corals of west Sumatra}, \textit{Science}, \textit{322}(5908), 1674--1678.

\bibitem[{\textit{Solomon and Sleep}(1974)}]{solomon1974some}
Solomon, S.~C., and N.~H. Sleep (1974), {Some simple physical models for
  absolute plate motions}, \textit{Journal of Geophysical Research},
  \textit{79}(17), 2557--2567.

\bibitem[{\textit{Suito and Hirahara}(1999)}]{suito1999simulation}
Suito, H., and K.~Hirahara (1999), {Simulation of postseismic deformations
  caused by the 1896 Riku-u Earthquake, northeast Japan: Re-evaluation of the
  viscosity in the upper mantle}, \textit{Geophysical Research Letters},
  \textit{26}(16), 2561--2564.

\bibitem[{\textit{Toda et~al.}(2011)\textit{Toda, Lin, and
  Stein}}]{toda2011using}
Toda, S., J.~Lin, and R.~S. Stein (2011), {Using the 2011 $M_W$ 9.0 off the
  Pacific coast of Tohoku Earthquake to test the Coulomb stress triggering
  hypothesis and to calculate faults brought closer to failure}, \textit{Earth,
  Planets and Space}, \textit{63}(7), 725--730.

\bibitem[{\textit{Tomita et~al.}(2015)\textit{Tomita, Kido, Osada, Hino, Ohta,
  and Iinuma}}]{tomita2015first}
Tomita, F., M.~Kido, Y.~Osada, R.~Hino, Y.~Ohta, and T.~Iinuma (2015), {First
  measurement of the displacement rate of the Pacific Plate near the Japan
  Trench after the 2011 Tohoku-Oki earthquake using GPS/acoustic technique},
  \textit{Geophysical Research Letters}, \textit{42}(20), 8391--8397.

\bibitem[{\textit{Uchida et~al.}(2016)\textit{Uchida, Asano, and
  Hasegawa}}]{uchida2016acceleration}
Uchida, N., Y.~Asano, and A.~Hasegawa (2016), {Acceleration of regional plate
  subduction beneath Kanto, Japan, after the 2011 Tohoku-oki earthquake},
  \textit{Geophysical Research Letters}, \textit{43}(17), 9002--9008.

\bibitem[{\textit{Ueda et~al.}(2003)\textit{Ueda, Ohtake, and
  Sato}}]{ueda2003postseismic}
Ueda, H., M.~Ohtake, and H.~Sato (2003), {Postseismic crustal deformation
  following the 1993 Hokkaido Nansei-oki earthquake, northern Japan: Evidence
  for a low-viscosity zone in the uppermost mantle}, \textit{Journal of
  Geophysical Research}, \textit{108}(B3).

\end{thebibliography}

\begin{figure}
\centering
\includegraphics[width=1.0\textwidth]{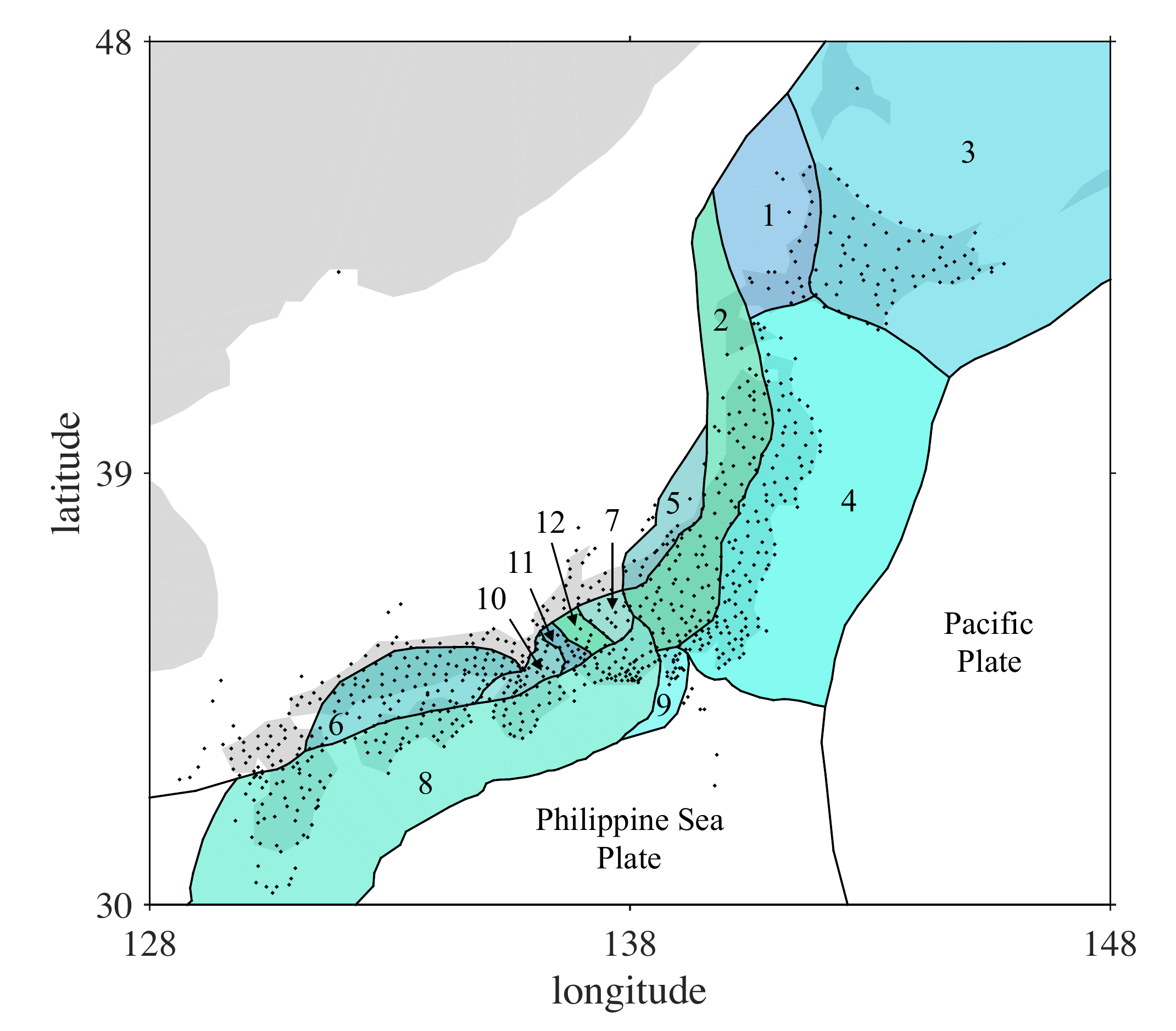}
\caption{Japan block model geometry \citep{loveless2016two} and GPS station locations (black circles). The numbers indicated here correspond to upper plate blocks referenced in Figure~\ref{JapanDeltaS}. Shading is distinct for each block considered in this study.}
\label{JapanMapAndBlocks}
\end{figure}

\begin{figure}
\centering
\includegraphics[width=1.0\textwidth]{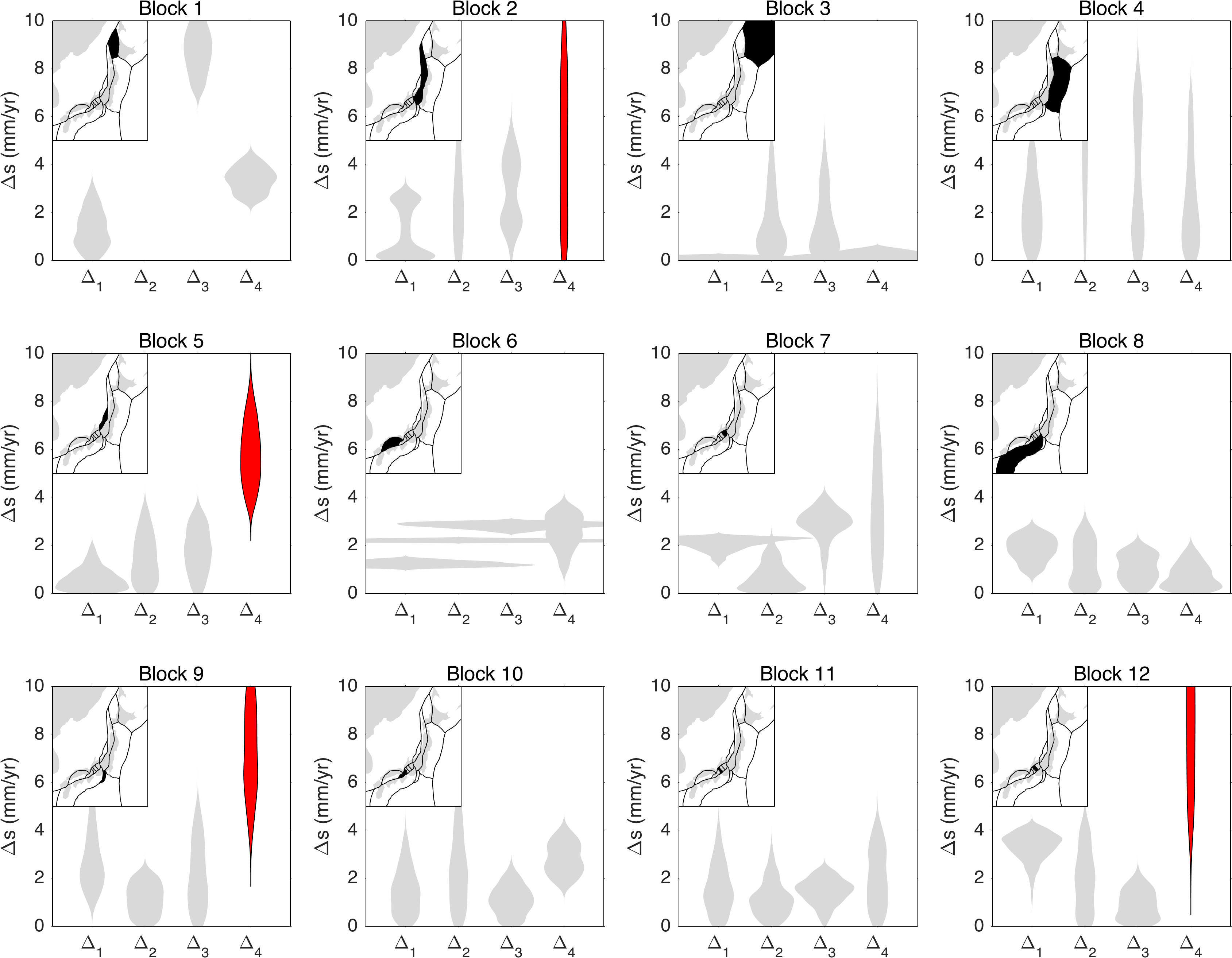}
\caption{Epoch-to-epoch speed changes, $\Delta s_i = | \langle s(E_{i+1}) \rangle - \langle s(E_{i}) \rangle |$, for the Japan crustal blocks shown in Figure~\ref{JapanMapAndBlocks}. Five, $\sim3.75$-year epochs are defined over the	interval April 1996-December 2015 (Table~\ref{table2}). The probability density functions show the range of plate speed changes and are shaded red if the change detection index satisfies three threshold criteria and gray otherwise. Inset figures show the greater Japan region with the current block shaded black. Epoch-to-epoch speed changes satisfy the change detection index in only the $\Delta_4$ interval (bracketing the Tohoku earthquake) for blocks 2, 5, 9, and 12 in central Honshu.}
\label{JapanDeltaS}
\end{figure}

\begin{table}[t]
\caption{Upper mantle viscosity (assuming a Maxwell viscoelastic rheology) estimates in Japan from studies of both post-seismic	and	post-glacial rebound. Viscosity	estimates from post-glacial rebound studies are $1.5-80$ times larger than those from post-seismic studies.}
\centering
\begin{tabular}{llll}
\hline
Study & Viscosity (Pa$\cdot$s) & Model type & Data type \\ \hline
\textit{Nur and Mavko} [1974] & $5.0\times10^{18}$ & post-seismic & coastlines \\
\textit{Nakada} [1986] & $4.0\times10^{20}$ & post-glacial & coastlines \\
\textit{Suito and Hirara} [1990] & $9.3\times10^{18}$ & post-seismic & leveling \\
\textit{Nakada et al.} [1991] & $2.0\times10^{19}$ & post-glacial & coastlines \\
\textit{Ueda et al.} [2003] & $4.0\times10^{18}$ & post-seismic & GPS \\
\textit{Nyst et al.} [2006] & $1.4\times10^{19}$ & post-seismic & GPS \\
\textit{Mavrommatis et al.} [2014] & $5.0\times10^{18}$ & post-seismic & GPS \\
\hline
\end{tabular}
\label{table1}
\end{table}

\begin{table}[t]
\caption{Temporal decomposition of GPS displacement time series into $\sim 3.75$-year long epochs \citep{loveless2016two}. Epochs $E_2$ and $E_4$ end the day before the	Tokachi	and Tohoku earthquakes respectively.}
\centering
\begin{tabular}{llll}
\hline
Epoch & Start date & End date & Duration (years) \\ \hline
$E_1$ & April 1, 1996 & December 31, 1999 & 3.75 \\
$E_2$ & January 1, 2000 & September 24, 2003 & 3.73 \\
$E_3$ & September 25, 2003 & June 30, 2007 & 3.77 \\
$E_4$ & July 1, 2007 & March 10, 2011 & 3.69 \\
$E_5$ & March 11, 2011 & December 31, 2014 & 3.81 \\ \hline
\end{tabular}
\label{table2}
\end{table}

\end{document}